\pdfoutput=1

\documentclass[11pt,twoside,a4paper,cmspaper,final,collab]{cms-tdr}
\def\svnVersion{484896P}\def\cmsCernNoTag{CERN-EP-2017-337}\def\cmsCernDate{\today}\def\cmsMessage{Published in Physical Review Letters as \href{http://dx.doi.org/10.1103/PhysRevLett.121.242301}{\doi{10.1103/PhysRevLett.121.242301.}}}

\begin{document}\cmsNoteHeader{HIN-16-014}

\hyphenation{had-ron-i-za-tion}
\hyphenation{cal-or-i-me-ter}
\hyphenation{de-vices}
\RCS$Revision: 484890 $
\RCS$HeadURL: svn+ssh://svn.cern.ch/reps/tdr2/papers/HIN-16-014/trunk/HIN-16-014.tex $
\RCS$Id: HIN-16-014.tex 484890 2018-12-14 22:54:17Z katatar $
\ifthenelse{\boolean{cms@external}}{\providecommand{\cmsSpar}[1]{#1}}{\providecommand{\cmsSpar}[1]{\sloppypar{#1}}}
\newcommand {\rootsNN}  {\ensuremath{\sqrt{\smash[b]{s_{_{\mathrm{NN}}}}}}}
\newcommand {\deta}     {\ensuremath{\Delta\eta}\xspace}
\newcommand {\dphi}     {\ensuremath{\Delta\phi}\xspace}
\newcommand {\pp}    {\ensuremath{\Pp\Pp}\xspace}
\newcommand {\PbPb}  {\ensuremath{\text{PbPb}}\xspace}
\newcommand {\ak}    {anti-\kt\xspace}
\newcommand {\ptg}    {\ensuremath{\pt^\gamma}\xspace}
\newcommand {\ptgvec}    {\ensuremath{\vec{p}_{\mathrm{T}}^\gamma}\xspace}
\newcommand {\ptj}    {\ensuremath{\pt^\text{jet}}\xspace}
\newcommand {\pgvec}    {\ensuremath{\vec{p}^\gamma}\xspace}
\newcommand {\pjetvec}    {\ensuremath{\vec{p}^{\kern1pt\text{jet}}}\xspace}
\newcommand {\ptrkvec}    {\ensuremath{\vec{p}^{\kern1pt\text{trk}}}\xspace}
\newcommand {\ptt}    {\ensuremath{\pt^\text{trk}}\xspace}
\newcommand {\pttvec}    {\ensuremath{\vec{p}_{\mathrm{T}}^{\kern1pt\text{trk}}}\xspace}
\newcommand {\etag}    {\ensuremath{\eta^\gamma}\xspace}
\newcommand {\etaj}    {\ensuremath{\eta^\text{jet}}\xspace}
\newcommand {\phig}    {\ensuremath{\phi^\gamma}\xspace}
\newcommand {\phij}    {\ensuremath{\phi^\text{jet}}\xspace}
\newcommand {\xigamma}    {\ensuremath{\xi^\gamma_{\mathrm{T}}}\xspace}
\newcommand {\xijet}    {\ensuremath{\xi^\text{jet}}\xspace}
\providecommand{\ptg}{\ensuremath{p_{\mathrm{T},\gamma}}\xspace}
\providecommand{\dphijg}{\ensuremath{\Delta\phi_{j\gamma}}\xspace}
\newcommand {\ntrk}  {\ensuremath{N^{\text{trk}}}\xspace}
\newcommand {\njet}  {\ensuremath{N^{\text{jet}}}\xspace}
\providecommand{\HYDJET}{\textsc{hydjet}\xspace}

\cmsNoteHeader{HIN-16-014}

\title{Observation of medium induced modifications of jet fragmentation in \PbPb collisions at \texorpdfstring{$\sqrtsNN=5.02$\TeV}{sqrt(s[NN]) = 5.02 TeV} using isolated-photon-tagged jets}

\date{\today}

\abstract{Measurements of fragmentation functions for jets associated with an isolated photon are presented for the first time in \pp and \PbPb collisions. The analysis uses data collected with the CMS detector at the CERN LHC at a nucleon-nucleon center-of-mass energy of 5.02\TeV. Fragmentation functions are obtained for jets with $\ptj > 30\GeVc$ in events containing an isolated photon with $\ptg> 60\GeVc$, using charged tracks with transverse momentum $\ptt > 1\GeVc$ in a cone around the jet axis. The association with an isolated photon constrains the initial \pt and azimuthal angle of the parton whose shower produced the jet. For central \PbPb collisions, modifications of the jet fragmentation functions are observed when compared to those measured in \pp collisions, while no significant differences are found in the 50\% most peripheral collisions. Jets in central \PbPb events show an excess (depletion) of low (high) \pt particles, with a transition around 3\GeVc. This measurement shows for the first time the in-medium shower modifications of partons (quark dominated) with well defined initial kinematics. It constitutes a new well-controlled reference for testing theoretical models of the parton passage through the QGP.}

\hypersetup{%
pdfauthor={CMS Collaboration},%
pdftitle={Observation of medium induced modifications of jet fragmentation in \PbPb collisions using isolated-photon-tagged jets},%
pdfsubject={CMS},%
pdfkeywords={CMS, physics, software, computing}}

\maketitle

A deconfined state of quarks and gluons, called the quark-gluon plasma (QGP)~\cite{Karsch:1995sy}, is believed to be produced on a short timescale in high energy nucleus-nucleus collisions~\cite{PhysRevD.27.140}. Occasionally, a pair of partons (quarks or gluons) in the colliding nuclei undergoes a high transverse momentum (\pt) scattering, a process that occurs over a shorter timescale. As they pass through and interact with the QGP, the scattered partons lose some of their energy~\cite{Appel:1985dq,Blaizot:1986ma,Gyulassy:1990ye,Wang:1991xy,Baier:1996sk,Zakharov:1997uu}. The relative importance of the various mechanisms by which these partons lose energy to the medium has been a continuous focus of the field of relativistic heavy ion collisions~\cite{Gyulassy:2000er, Djordjevic:2003zk, Ovanesyan:2011xy, Wang:2001ifa, Arnold:2002ja, Burke:2013yra}.

The outgoing hard-scattered partons eventually fragment, and each forms a jet of collimated particles. The CERN LHC collaborations have conducted various (di)jet studies: modifications of the jet yield in the medium (jet quenching)~\cite{Aad:2014bxa, Adam:2015ewa, Khachatryan:2016jfl}, jet fragmentation functions (the probability for a parton to fragment into particles carrying a given fraction of the jet momentum)~\cite{Chatrchyan:2014ava, Aaboud:2017bzv}, missing \pt in dijet systems~\cite{Aad:2010bu, Chatrchyan:2011sx, Khachatryan:2015lha}, jet-track correlations~\cite{Khachatryan:2016tfj}, and the radial \pt profile of tracks within jets~\cite{Chatrchyan:2013kwa}. However, in these analyses, the energy lost by the partons diminishes the information about their initial properties. One way to overcome this challenge is to study processes in which the initial hard scattering produces a parton and an electroweak boson: the bosons do not experience quantum chromodynamic interactions and are largely unaffected by the QGP. At leading order, bosons are produced back-to-back with an associated parton having close to the same \pt, modulo secondary effects such as multiple scatterings of the initial partons or initial state radiation. As a result, the jets associated with the boson should have parent partons whose \pt before any energy loss occurs is well defined. In addition, at LHC energies, the electroweak-boson+jet production is dominated by quark jets for $\ptj>30$\GeVc~\cite{Neufeld:2010fj, Kang:2017xnc, Dai:2012am}, therefore providing information specifically on quark energy loss. 

The CMS Collaboration measured the azimuthal correlation and momentum imbalance of isolated-photon+jet pairs in \pp and \PbPb collisions at nucleon-nucleon center-of-mass energies of $\rootsNN = 2.76$ and 5.02\TeV~\cite{Chatrchyan:2012gt, Sirunyan:2017qhf} and of Z+jet pairs at $\rootsNN = 5.02\TeV$~\cite{Sirunyan:2017jic}. In related studies, experiments at RHIC extracted jet fragmentation functions associated with photons without fully reconstructing the jets, but rather by studying direct-photon+hadron correlations~\cite{Adare:2012qi, STAR:2016jdz}. This Letter presents the first measurement of the fragmentation function of jets that are fully reconstructed and associated with an isolated photon (\ie, one with no significant energy deposited around its location in the detector). This definition suppresses dijet events in which a high-\pt photon originates from one of the jets, either via collinear fragmentation of a parton (``fragmentation photons") or via decays of neutral mesons (``decay photons"). The analysis uses \PbPb and \pp data at $\rootsNN = 5.02\TeV$ collected in 2015 and corresponding to integrated luminosities of 404\mubinv and 27.4\pbinv, respectively. 

\cmsSpar{
Photon-tagged fragmentation
functions are presented as distributions of $\xijet = \ln{\bigl[\abs{\pjetvec}^{2}/(\ptrkvec \cdot \pjetvec)\bigr]}$, where \pjetvec and \ptrkvec are the 3-momenta of the jet and charged particle, respectively, and $\xigamma = \ln{\bigl[-\abs{\ptgvec}^{2}/(\pttvec \cdot \ptgvec)\bigr]}$, where \ptgvec and \pttvec are the \pt with respect to the beam direction of the photon and charged-particle, respectively. The $\xijet$ variable gives the fragmentation pattern with respect to \pt of the reconstructed jet~\cite{Arleo:2004xj, Casalderrey-Solana:2015vaa}, and can be compared directly with results obtained using a dijet sample~\cite{Chatrchyan:2014ava}. The $\xigamma$ variable is used to characterize the fragmentation pattern with respect to the \pt of the initial parton before any energy loss occurred. The \ptgvec is used instead of \pgvec because the photon and parton from a hard scattering have the same $\abs{\pt}$ at leading order, but not necessarily the same magnitude of longitudinal momentum.}

The central feature of the CMS detector is a superconducting solenoid, providing a magnetic field of 3.8\unit{T}. Within the solenoid volume are a pixel and strip tracker, an electromagnetic calorimeter (ECAL), and a hadron calorimeter (HCAL). Hadron forward (HF) calorimeters extend the pseudorapidity coverage up to $\abs{\eta}=5.2$. In the case of \PbPb events, the HF signals are used to determine the degree of overlap (``centrality'') of the two colliding Pb nuclei~\cite{Chatrchyan:2011sx} and the event-by-event azimuthal angle of maximum particle density (``event plane'')~\cite{Chatrchyan:2012xq}. A more detailed description of the CMS detector can be found in Ref.~\cite{Chatrchyan:2008zzk}.

The event samples are selected online with a dedicated photon trigger requiring one photon with $\ptg>40$\GeVc~\cite{Sirunyan:2017qhf}, and are subjected to offline requirements to remove non-collision events~\cite{Khachatryan:2016odn,Sirunyan:2017qhf}. For jets and photons, the reconstruction algorithms, analysis selections, and corrections for the energy scale and resolution are the same as in Ref.~\cite{Sirunyan:2017qhf}. For the analysis of \PbPb collisions, the event centrality is defined as the fraction of the total inelastic hadronic cross section, starting at 0\% for the most central collisions, and is evaluated as percentiles of the distribution of the energy deposited in the HF calorimeters~\cite{Chatrchyan:2011sx}. Results are presented in four centrality intervals, ranging from a central 0--10\% (\ie, the 10\%  of the events having the largest overlap area of the two nuclei~\cite{Miller:2007ri}), to a peripheral 50--100\% (the one closest to a \pp-like environment) intervals.

The photon candidates are restricted to the barrel region of the ECAL, $\abs{\etag} < 1.44$, and are required to have $\ptg > 60\GeVc$. Electron contamination
and anomalous signals caused by the interaction of heavily ionizing particles with the silicon avalanche photodiodes used for the ECAL readout are removed~\cite{Chatrchyan:2012vq}. Background from ECAL showers induced by hadrons are rejected using the ratio of HCAL over ECAL energy inside a cone of radius $\Delta R = \sqrt{\smash[b]{(\deta)^2 + (\dphi)^2}}  = 0.15$ around the photon candidate~\cite{Khachatryan:2010fm,Chatrchyan:2012vq}. Background contributions from fragmentation and decay photons are rejected by imposing isolation requirements~\cite{Sirunyan:2017qhf,Chatrchyan:2012vq}. The dominant remaining background is ECAL showers initiated by isolated neutral mesons decaying into pairs of photons that are reconstructed as a single photon. Because the pattern of energy deposited in the ECAL (\ie, the ``shower shape") is wider in $\eta$ for these decay photons, their contribution can be reduced by a factor of $\sim$2 using an upper limit on the width of the $\eta$ distribution~\cite{Chatrchyan:2012vq}.

The energy of the reconstructed photons is corrected to account for the effects of the material in front of the ECAL and for incomplete containment of the shower energy in the ECAL crystals~\cite{Khachatryan:2015iwa}. An additional correction is applied in \PbPb collisions to account for the contribution of the underlying event (UE). The corrections are obtained from photon events simulated using the CUETP8M1 tune~\cite{Khachatryan:2015pea} of the \PYTHIA~8.212~\cite{Sjostrand:2014zea} Monte Carlo (MC) event generator. The effect of the \PbPb UE is modeled by embedding the \PYTHIA output in events generated using \HYDJET 1.9~\cite{Lokhtin:2005px}. The background simulation is tuned to reproduce the observed charged-particle multiplicity and \pt spectrum in \PbPb data. The size of the resulting energy correction for isolated photons varies from 0 to 10\%, depending on \ptg and the centrality of the collision.

The particle-flow (PF) algorithm~\cite{Sirunyan:2017ulk} is used for the jet reconstruction with the \ak algorithm provided in the \FASTJET framework~\cite{Cacciari:2008gp,Cacciari:2011ma} with a distance parameter $R = 0.3$. In order to subtract the UE background in \PbPb collisions, an iterative algorithm~\cite{Kodolova:2007hd} is employed~\cite{Chatrchyan:2011sx,Chatrchyan:2012nia,Chatrchyan:2012gt}. In \pp collisions, where the UE level is negligible, jets are reconstructed without UE subtraction (for a jet with $\pt=30$\GeVc, the UE contribution is at most 1\%). The jet energy corrections are derived from simulation, separately for \pp and \PbPb, and are confirmed via energy-balance methods in \pp data~\cite{Khachatryan:2016kdb}. Jets with $\abs{\etaj} < 1.6$ and corrected $\ptj > 30\GeVc$ are selected. In order to compare the \PbPb and \pp results, the jet energy and \phij in \pp events are smeared to match the corresponding resolutions in each of the \PbPb centrality intervals. The parametrization of the energy smearing function is given in Ref.~\cite{Sirunyan:2017qhf}. To match the 0--10\% \PbPb data, the energy resolution of \pp jets with $\ptj=30(90)$\GeVc changes from 18\%\,(12\%) to 35\%\,(17\%). The change in angular resolution is negligible ($<$2.2\%).

In each event, photon+jet pairs are formed by associating the highest-\ptg
isolated photon candidate with all jets that pass the jet selection criteria. An azimuthal separation of $\dphijg = \abs{\phij - \phig} > 7\pi/8$ is applied to the photon+jet pairs to suppress contributions from background jets (not from the same hard scattering as the photon) and photon+multijet events (from an early splitting of the original parton). The tracks used in the measurement of the fragmentation function have $\ptt>1$\GeVc and are required to fall within a cone of radius $\Delta R = 0.3$ around the jet direction. These selection criteria, as well as the corrections for tracking efficiency, detector acceptance, and misreconstruction rate, are the same as in Ref.~\cite{Khachatryan:2016odn}.

The selected charged-particle tracks (\ntrk) are used in conjunction with the selected photon+jet pairs to determine the fragmentation functions, $({1}/{\njet})\,({\rd\ntrk}/{\rd\xijet})$ and $({1}/{\njet})\,({\rd\ntrk}/{\rd\xigamma})$, where \njet represents the total number of photon+jet pairs. To isolate the contribution of photons, jets, and charged particles that are produced in the same hard scattering in \PbPb collisions, several combinatorial backgrounds are subtracted: tracks from the UE that fall within the cone around the selected jet, misidentified jets resulting from UE fluctuations, and jets not produced in the same hard parton-parton scatterings as the photon. The shape and magnitude of these contributions are estimated from data with an event mixing procedure, in which either the isolated photon or the jet is combined with jets and tracks found in events chosen randomly from a minimum-bias (MB) \PbPb dataset with similar event characteristics (centrality, interaction vertex position, and event plane angle). The background contribution from UE tracks is estimated by correlating each selected jet with tracks from MB events. The backgrounds from jets produced by UE fluctuations or a different parton-parton scattering are estimated by correlating each selected photon with jets from randomly selected MB events as in Refs.~\cite{Chatrchyan:2012gt,Sirunyan:2017qhf}. The normalizations of these combinatorial background fragmentation functions are given by the number of MB events used. Simulations predict that the average UE particle density is slightly different between a MB event and an event containing a hard scattering, even when the two have the same collision centrality. Therefore, the normalized background distributions are further scaled with a residual factor to account for this effect. The final correction accounts for the photon purity, defined to be the fraction of non-decay photons within the collection of isolated photon candidates that pass the shower shape requirement. This fraction is extracted from the data using a template fit to the shower shape distribution, and is $\sim$65\% (85\%) in 0--10\% (0--50\%) PbPb collisions~\cite{Sirunyan:2017qhf,Chatrchyan:2012gt}. The shape of the fragmentation functions from decay photons is estimated by repeating the analysis selecting photons with wider shower shapes. 

Several sources of systematic uncertainty are considered: photon purity, energy scale, and isolation, electron contamination, jet energy scale and resolution, tracking efficiency, and UE background. The total uncertainty is the sum in quadrature of the individual uncertainties. The quoted systematic uncertainties are an average over all \xijet and \xigamma bins and, in the case of the \PbPb results, are quoted only for events with 0--10\% centrality. To evaluate the systematic uncertainties related to the isolated photons, the same procedures as in Ref.~\cite{Sirunyan:2017qhf} are applied. The uncertainty in the photon purity estimation is evaluated by varying the components of the shower shape template~\cite{Chatrchyan:2012gt}. The maximum variations with respect to the nominal case are propagated as systematic uncertainties, amounting to 2.8\,(5.4)\% for the \PbPb events, and 0.4\,(0.4)\% for the \pp results, for \xijet\,(\xigamma). In the following, the uncertainties will continue to be quoted for central \PbPb first, then for \pp. The total systematic uncertainties resulting from the experimental criteria for an isolated photon are 1.7\,(1.1)\% and 0.9\,(0.7)\% for \xijet\,(\xigamma). The residual data-to-simulation photon energy scale difference after applying the photon energy corrections is also quoted as a systematic uncertainty of 1.2\,(1.2)\% and $<$0.1\% for \xijet\,(\xigamma). An uncertainty for the level of electron contamination is evaluated by repeating the analysis without applying the electron rejection criteria, and scaling down the difference in the $\xi$ observables to the remaining electron level of contamination after applying the electron rejection, giving 0.6\,(0.5)\% and $<$0.1\,(0.1)\% for \xijet\,(\xigamma). The efficiency for selecting photons has been extracted from MC calculations as a function of photon \pt. An uncertainty is assigned by comparing the results to the ones obtained with a correction derived after loosening the selection criteria, giving 0.1\,(0.5)\% and $<$0.1\,($<$0.1)\% for \xijet\,(\xigamma). The uncertainty related to the jet energy scale~\cite{Sirunyan:2017qhf} amounts to 7.3\,(6.5)\% and 2.4\,(0.6)\% for \xijet\,(\xigamma), while the energy resolution~\cite{Sirunyan:2017qhf} gives uncertainties of 2.8\,(1.7)\% and 0.7\,(0.5)\%. A systematic uncertainty is assigned to account for long-range $\eta$ correlations~\cite{Khachatryan:2016tfj} that contribute to the UE. It is estimated by constructing the observables using tracks lying within the same azimuthal angle as the jet, but separated by a large pseudorapidity interval, $1.5 < \deta < 2.4$. The uncertainty is found to be 4.1\,(3.3)\% and 1.7\,(1.5)\% for \xijet\,(\xigamma). The uncertainty related to the tracking inefficiency is estimated as the difference in the track reconstruction efficiency between data and simulation~\cite{Khachatryan:2016odn}. It is 5\,(4)\% for \PbPb, \pp) data, for both \xijet  and \xigamma, and is constant as a function of the track \pt and event centrality.

For the \PbPb results only, two additional sources of systematic uncertainties are considered. One accounts for possible inaccuracies in the background subtraction by combining two independent components. First, the effect of the background subtraction method is estimated using an alternative procedure (the so-called $\eta$-reflection method~\cite{Chatrchyan:2014ava}), which has different sensitivities to various background sources. Second, results are compared to the ones where background distributions are not scaled for the UE particle density difference seen when comparing simulated MB events with those containing a hard scattering. The combined difference of 3.6\,(3.5)\% for \xijet(\xigamma) is assigned as the uncertainty. The second uncertainty accounts for differences in the jet energy response due to \xijet and \xigamma variances in the jet fragmentation pattern, as studied in simulation~\cite{Sirunyan:2017qhf}. The observed differences of 11\% for $\xijet < 1$ and 4.3\,(7.0)\% for $\xijet\,(\xigamma) > 2.5$ are propagated as uncertainties in the corresponding \xijet and \xigamma regions for the \PbPb results, while there are no significant differences in other $\xijet$ and $\xigamma$ bins.

\begin{figure}[tbp]
 \centering
   \includegraphics[width=0.49\textwidth]{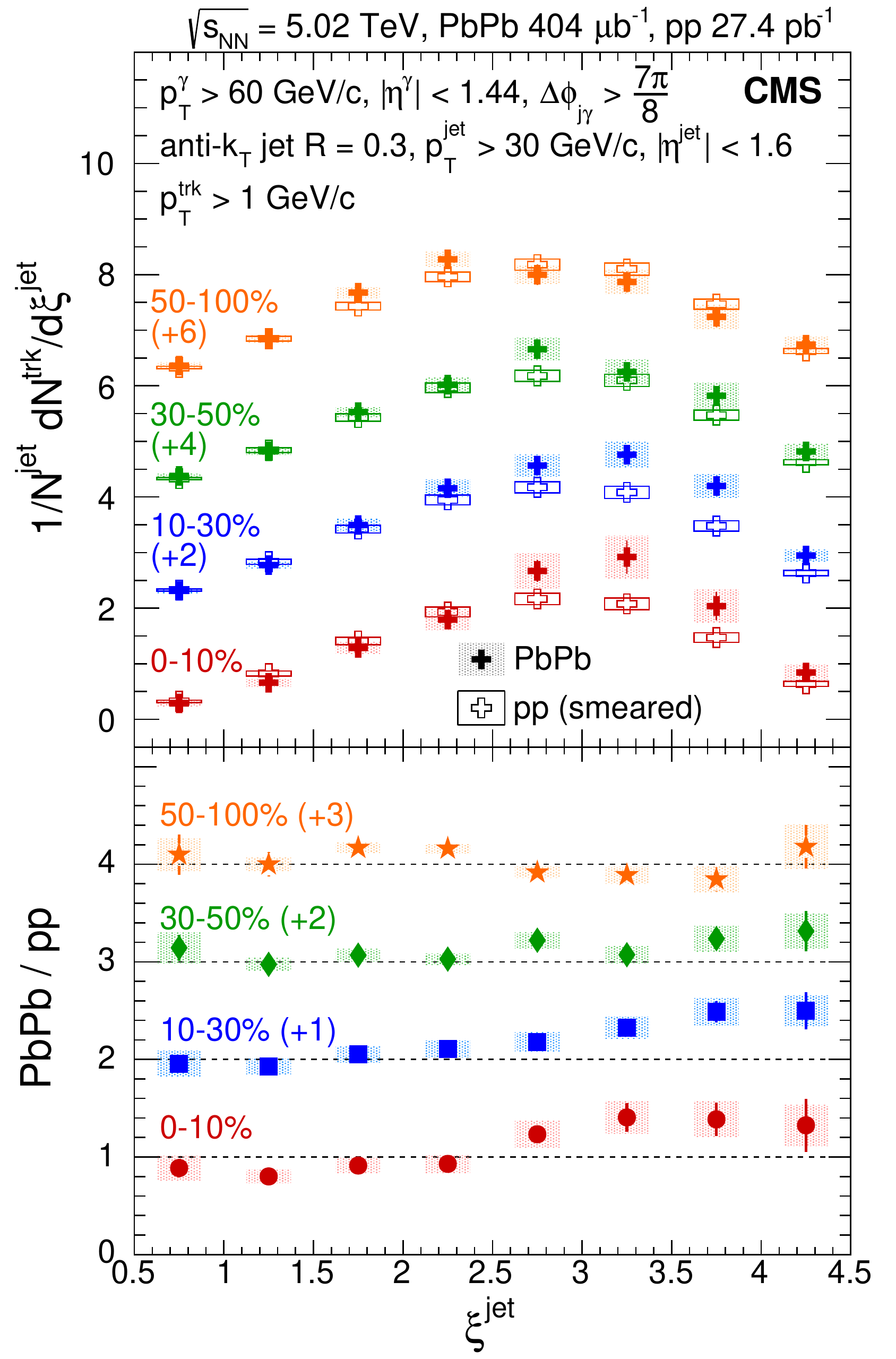}
   \caption{
     {Top:} The centrality dependence of the $\xijet$ distribution for jets associated with an isolated photon for \PbPb (full crosses) and \pp (open crosses) collisions. The \pp results are smeared for each \PbPb centrality bin, and data for each centrality bin are shifted vertically as indicated. {Bottom:} The ratios of the \PbPb over smeared \pp distributions. The vertical bars through the points represent statistical uncertainties, while the colored boxes indicate systematic uncertainties.
     }
     \label{fig:xijet}
\end{figure}

Figure~\ref{fig:xijet} shows the photon-tagged fragmentation functions as a
function of $\xijet$ for both \PbPb and \pp collisions, together with their ratio of the \PbPb to
pp results. The $\xijet$ distributions in \PbPb collisions
represent the fragmentation pattern of jets that may have lost energy through
interactions with the medium, while those for \pp
stand for unquenched jets. Because of this possibility of quenching, the collections of reconstructed jets in \PbPb and \pp collisions do not necessarily have the same \ptj spectrum, even though the samples are selected based on photon energy. The $\xijet$ distributions for 50--100\% centrality \PbPb collisions are consistent with those in \pp collisions. In more central collisions, an enhancement of the fragmentation function in \PbPb collisions with respect to
the reference \pp data is observed for $\xijet > 2.5$ (corresponding to $\ptt \lesssim 2.5$\GeVc for $\ptj = 30$\GeVc and $\Delta R = 0$ between the track and the jet), indicating that there is
a small excess of soft particles near the jet. Additionally, a slight
suppression of the fragmentation function in the region $0.5 < \xijet < 2.5$ (corresponding to $18\gtrsim \ptt \gtrsim 2.5$\GeVc for $\ptj = 30$\GeVc and $\Delta R = 0$) is also observed in the most central \PbPb collisions.

\begin{figure}[tbp]
 \centering
   \includegraphics[width=0.49\textwidth]{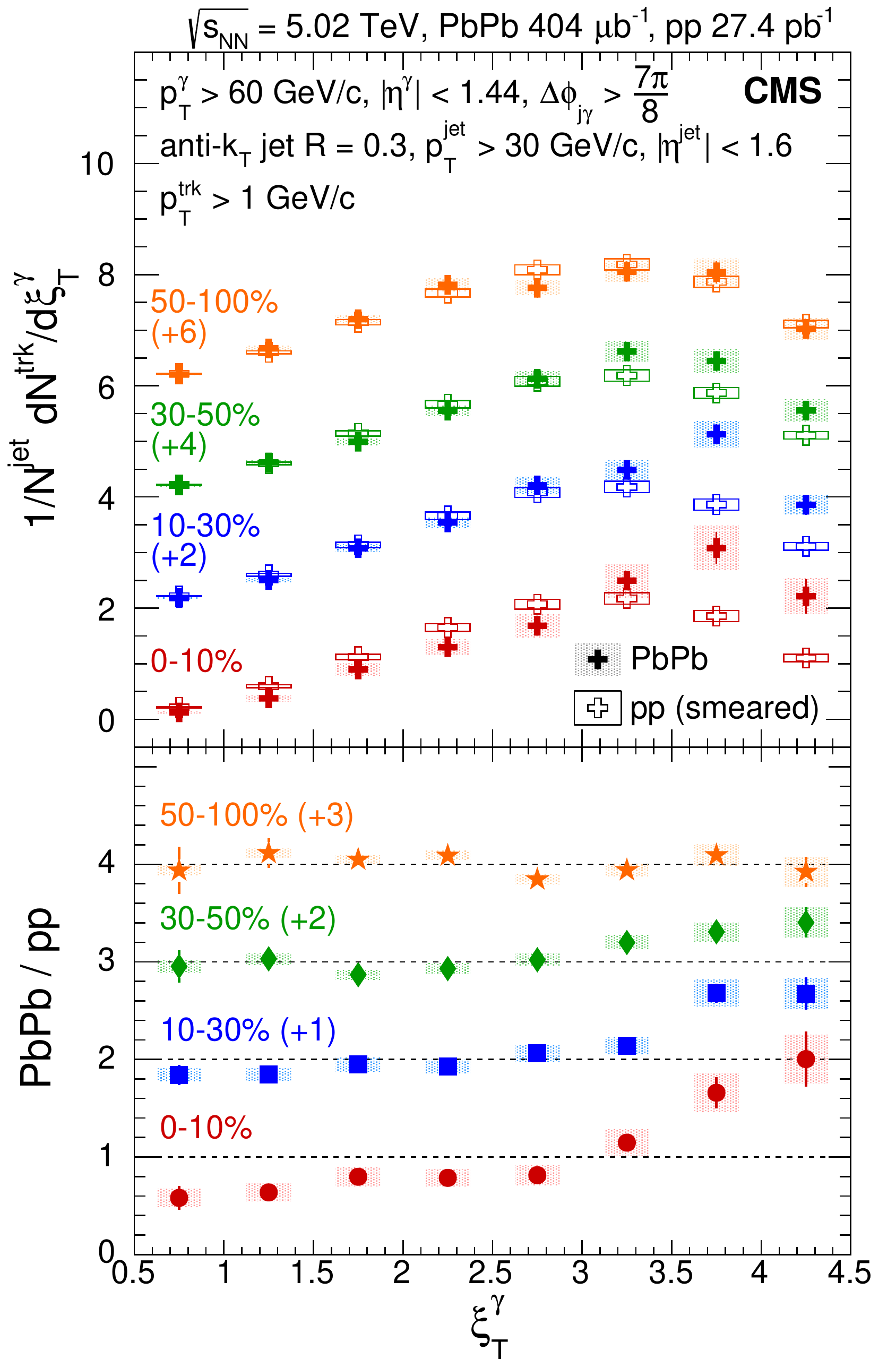}
   \caption{
       {Top:} The centrality dependence of the $\xigamma$ distribution for jets associated with an isolated photon for \PbPb (full crosses) and \pp (open crosses) collisions. The \pp results are smeared for each \PbPb centrality bin, and data for each centrality bin are shifted vertically as indicated. {Bottom:} The ratios of the \PbPb over smeared \pp distributions. The vertical bars through the points represent statistical uncertainties, while the colored boxes indicate systematic uncertainties.
     }
     \label{fig:xigamma}
\end{figure}

Figure~\ref{fig:xigamma} shows the photon-tagged fragmentation functions as a
function of $\xigamma$ for \PbPb and \pp collisions, as well as their ratio. As for \xijet, the \xigamma distributions for peripheral \PbPb events are consistent with those in \pp data. In more central collisions, an enhancement is observed in the \PbPb data relative to \pp data in the $\xigamma > 3$ region (corresponding to $\ptt \lesssim 3$\GeVc for $\ptg = 60$\GeVc and $\Delta\phi = \pi$ between the track and the photon). The magnitude of this enhancement increases as the \PbPb collisions become more central, and is more significant than the enhancement observed in the \xijet distributions. The differences between the \pp and \PbPb distributions are quantified by comparing the two distributions using a $\chi^2$ test. The $p$-values from the test are $10^{-3}$  and $10^{-20}$ for the \xijet and \xigamma distributions, respectively. Similarly, the \xijet distribution was compared to that for \xigamma within the same system. The $p$-values from the test are $10^{-3}$  and $10^{-10}$ for the \PbPb and \pp results, respectively. A suppression of the \xigamma distribution for the most central \PbPb collisions is also observed for $ 0.5 < \xigamma < 3$ (corresponding to $36 \gtrsim \ptt \gtrsim 3$\GeVc for $\ptg = 60$\GeVc and $\Delta\phi = \pi$). This pattern of suppression and enhancement is direct evidence for energy loss by high-\pt partons as they traverse the high-density medium created in heavy ion collisions~\cite{Wang:1996yh,Wang:1996pe}.

The enhancement at large \xijet and \xigamma, together with the suppression at lower values, indicates that the showers of partons that emerge from the medium contain more lower-energy particles in \PbPb collisions. These particles can originate directly from high energy partons that lose energy, as well as from the medium response, the recoil of the medium as the parton traverses through~\cite{Casalderrey-Solana:2015vaa, KunnawalkamElayavalli:2017hxo, Wang:2013cia}.
The enhancements and suppressions of the fragmentation patterns measured using a photon to estimate the initial parton energy (\xigamma) are observed to be more pronounced than the ones measured using detector level jet energy (\xijet). Qualitatively, this is not surprising. Because of effects such as out-of-cone radiation not being captured in the \ak $R=0.3$ jet area, a shift of the distributions to lower values for \xijet compared to \xigamma is observed even in the \pp results. In \PbPb collisions, an additional shift happens when the jet is quenched: the jet momentum becomes lower than that of the parton, resulting in a shift to lower \xijet. As a result, the modifications in the ratio \PbPb/pp are weakened by quenching in the \xijet case. This analysis, using events selected with a photon trigger which ensures that the initial parton \pt spectra are the same for \pp and \PbPb, will allow testing the theoretical modelling of both the parton shower modification in the medium and the medium response to the passage of that parton.

In summary, the fragmentation functions of jets associated with isolated photons are measured for the first time in \pp and \PbPb data collected at $\rootsNN = 5.02\TeV$ by CMS. Fragmentation patterns as functions of $\xijet = \ln{\bigl[\abs{\pjetvec}^{2}/(\ptrkvec\cdot \pjetvec)\bigr]}$ and $\xigamma = \ln{\bigl[-\abs{\ptgvec}^{2}/(\pttvec \cdot \ptgvec)\bigr]}$ are constructed using charged particles with $\ptt > 1\GeVc$, for jets with $\ptj > 30\GeVc$ tagged by an isolated photon with $\ptg > 60\GeVc$. When compared to the \pp results, the \xijet and \xigamma distributions in central \PbPb collisions show an excess of low-\pt particles and a depletion of high-\pt particles inside the jet. This observation is more apparent in the \xigamma distributions, where the photon-based selection allows for tagging the properties of the initial parton before quenching occurs. This measurement shows for the first time the in-medium parton shower modifications for events with well-defined initial parton kinematics, and constitutes a new well-controlled reference for testing theoretical models of the parton's passage through the QGP.

\begin{acknowledgments}
We congratulate our colleagues in the CERN accelerator departments for the excellent performance of the LHC and thank the technical and administrative staffs at CERN and at other CMS institutes for their contributions to the success of the CMS effort. In addition, we gratefully acknowledge the computing centers and personnel of the Worldwide LHC Computing Grid for delivering so effectively the computing infrastructure essential to our analyses. Finally, we acknowledge the enduring support for the construction and operation of the LHC and the CMS detector provided by the following funding agencies: BMWFW and FWF (Austria); FNRS and FWO (Belgium); CNPq, CAPES, FAPERJ, and FAPESP (Brazil); MES (Bulgaria); CERN; CAS, MoST, and NSFC (China); COLCIENCIAS (Colombia); MSES and CSF (Croatia); RPF (Cyprus); SENESCYT (Ecuador); MoER, ERC IUT, and ERDF (Estonia); Academy of Finland, MEC, and HIP (Finland); CEA and CNRS/IN2P3 (France); BMBF, DFG, and HGF (Germany); GSRT (Greece); OTKA and NIH (Hungary); DAE and DST (India); IPM (Iran); SFI (Ireland); INFN (Italy); MSIP and NRF (Republic of Korea); LAS (Lithuania); MOE and UM (Malaysia); BUAP, CINVESTAV, CONACYT, LNS, SEP, and UASLP-FAI (Mexico); MBIE (New Zealand); PAEC (Pakistan); MSHE and NSC (Poland); FCT (Portugal); JINR (Dubna); MON, RosAtom, RAS, RFBR and RAEP (Russia); MESTD (Serbia); SEIDI, CPAN, PCTI and FEDER (Spain); Swiss Funding Agencies (Switzerland); MST (Taipei); ThEPCenter, IPST, STAR, and NSTDA (Thailand); TUBITAK and TAEK (Turkey); NASU and SFFR (Ukraine); STFC (United Kingdom); DOE and NSF (USA).
\end{acknowledgments}

\bibliography{auto_generated} 
\cleardoublepage \appendix\section{The CMS Collaboration \label{app:collab}}\begin{sloppypar}\hyphenpenalty=5000\widowpenalty=500\clubpenalty=5000\vskip\cmsinstskip
\textbf{Yerevan~Physics~Institute,~Yerevan,~Armenia}\\*[0pt]
A.M.~Sirunyan, A.~Tumasyan
\vskip\cmsinstskip
\textbf{Institut~f\"{u}r~Hochenergiephysik,~Wien,~Austria}\\*[0pt]
W.~Adam, F.~Ambrogi, E.~Asilar, T.~Bergauer, J.~Brandstetter, E.~Brondolin, M.~Dragicevic, J.~Er\"{o}, A.~Escalante~Del~Valle, M.~Flechl, M.~Friedl, R.~Fr\"{u}hwirth\cmsAuthorMark{1}, V.M.~Ghete, J.~Grossmann, J.~Hrubec, M.~Jeitler\cmsAuthorMark{1}, A.~K\"{o}nig, N.~Krammer, I.~Kr\"{a}tschmer, D.~Liko, T.~Madlener, I.~Mikulec, E.~Pree, N.~Rad, H.~Rohringer, J.~Schieck\cmsAuthorMark{1}, R.~Sch\"{o}fbeck, M.~Spanring, D.~Spitzbart, A.~Taurok, W.~Waltenberger, J.~Wittmann, C.-E.~Wulz\cmsAuthorMark{1}, M.~Zarucki
% ...
\vskip\cmsinstskip
% ...
1:~Also at~Vienna~University~of~Technology,~Vienna,~Austria\\
% ...
\end{sloppypar}
\end{document}